\documentclass[11pt]{article}
\usepackage{graphicx}
\def\vec#1{\ifmmode
\mathchoice{\mbox{\boldmath$\displaystyle\bf#1$}}
{\mbox{\boldmath$\textstyle\bf#1$}}
{\mbox{\boldmath$\scriptstyle\bf#1$}}
{\mbox{\boldmath$\scriptscriptstyle\bf#1$}}\else
{\mbox{\boldmath$\bf#1$}}\fi}

\begin{document}

\begin{flushleft}
\hspace{1 cm}  \\*[2 cm]
\end{flushleft}

\begin{center}
  {\LARGE Effect of Systematic Uncertainty Estimation \\*[0.2 cm]
    on the Muon $g-2$ Anomaly}
\end{center}

\vspace{1.5 cm}

\begin{center}
Glen Cowan      
\end{center}

\vspace{0.5 cm}

\begin{center}
Physics Department, Royal Holloway, University of London, 
Egham, TW20 0EX, U.K. 
\end{center}

\vspace{2 cm}

\begin{abstract}
  The statistical significance that characterizes a discrepancy
  between a measurement and theoretical prediction is usually
  calculated assuming that the statistical and systematic
  uncertainties are known.  Many types of systematic uncertainties
  are, however, estimated on the basis of approximate procedures and
  thus the values of the assigned errors are themselves uncertain.
  Here the impact of the uncertainty {\it on the assigned uncertainty}
  is investigated in the context of the muon $g-2$ anomaly.  The
  significance of the observed discrepancy between the Standard Model
  prediction of the muon's anomalous magnetic moment and measured
  values are shown to decrease substantially if the relative
  uncertainty in the uncertainty assigned to the Standard Model
  prediction exceeds around 30\%.  The reduction in sensitivity
  increases for higher significance, so that establishing a $5\sigma$
  effect will require not only small uncertainties but the
  uncertainties themselves must be estimated accurately to correspond
  to one standard deviation.
\end{abstract}

\vspace{2 cm}

\clearpage

\section{Introduction}

The recent measurement of the muon's anomalous magnetic moment at
Fermilab's Muon $g-2$ Experiment \cite{bib:FNAL2021}, when averaged
with the 2006 value from Brookhaven \cite{bib:BNL2006} was found to be
in disagreement with the Standard Model (SM) prediction \cite{bib:SMpred}
with a significance of $4.2 \sigma$.

The significance of the discrepancy treats both the measurements and
the theoretical prediction as Gaussian distributed quantities whose
standard deviations are exactly known.  Although the statistical
errors are no doubt estimated with negligible uncertainty, this is not
necessarily the case for systematic errors, particularly those of the
SM prediction.

Here the method for incorporating uncertainties on reported systematic
errors of Ref.~\cite{bib:Cowan2019} is applied to the muon $g-2$
significance. Section~\ref{sec:inputs} describes the input values for
the analysis and Sec.~\ref{sec:model} gives a brief description of the
statistical model.  Results are shown in Sec.~\ref{sec:results} and
finally some conclusions are drawn in Sec.~\ref{sec:conc}.

The recent result from Borsanyi et al. \cite{bib:Borsanyi2021} using
lattice QCD gives an SM prediction for $g-2$ that differs less from
the measured value.  As the purpose of this note is to illustrate the
importance of the assigned systematic uncertainty for the case of a
significant discrepancy, we focus on the $4.2\sigma$ difference
highlighted in Ref.~\cite{bib:FNAL2021} and leave the lattice result
for future consideration.

\section{Input values}
\label{sec:inputs}

To simplify the numerical treatment and presentation of the results,
the values $a_{\mu} = (g-2)/2$ are transformed according to

\begin{equation}
  \label{eq:ytrans}
  y = a_{\mu} \times 10^9 - 1165900 \;.
\end{equation}

\noindent The recent FNAL measurement \cite{bib:FNAL2021} when
combined with the 2006 value from BNL \cite{bib:BNL2006} results
in an averaged experimental value of

\[
y_{\rm exp} = 20.61 \pm 0.41 \;.
\]

\noindent The uncertainty reflects both statistical (0.37) and
systematic (0.17) errors.  The Standard Model prediction is given in
Ref.~\cite{bib:SMpred} as

\[
y_{\rm SM} = 18.10 \pm 0.43 \;.
\]

\noindent The uncertainty is dominated by the hadronic vacuum
polarization (0.40) and to a lesser extent the hadronic light-by-light
contribution (0.18).

\section{Including uncertainties in estimates of systematic errors}
\label{sec:model}

In this note, the gamma variance model of Ref.~\cite{bib:Cowan2019} is
used to include an uncertainty in assigned systematic errors into the
significance of the observed anomaly.  This model treats values $v$ of
systematic variances (i.e., $v = s^2$ where $s$ is the estimated
standard deviation) not as fixed constants but rather as estimates
that follow a gamma distribution.  The expectation values $E[v]$
become adjustable parameters of the model, and the standard deviations
$\sigma_v$ are fitted to reflect the accuracy with which the
systematic error is estimated.  To do this, the analyst supplies
parameters $r = \sigma_v / 2 E[v] \approx \sigma_s/E[s]$, which to
first approximation represent the relative uncertainty in the assigned
systematic errors.

Here we apply this method only to the uncertainty in the Standard
Model prediction $y_{\rm SM}$.  This is by far the largest systematic
uncertainty and owing to its theoretical origin it is inherently
difficult to estimate precisely.  That is, we treat the value of
$v_{\rm SM} = (0.43)^2$ as a gamma-distributed estimate of the true
variance $\sigma_{\rm SM}^2$ of $y_{\rm SM}$, and we assign to its
distribution a relative ``error on the error'' $r_{\rm SM}$.  In
principle, the model can be applied to any of the assigned
uncertainties, including the experimental systematic error.  As this
is substantially smaller than the SM uncertainty we leave this for
future investigation.

The likelihood function of the gamma variance model contains as free
parameters the mean $\mu$ and variance $\sigma_{\rm SM}^2$ of $y_{\rm
  SM}$.  The profile log-likelihood function is obtained by evaluating
$\sigma_{\rm SM}^2$ with the value that maximizes the likelihood for a
given $\mu$.  This is up to an additive constant a function $Q(\mu)$
that plays the role of the $\chi^2$ in a least-squares average, but
with the usual quadratic term for $y_{\rm SM}$ replaced by a
logarithmic one:

\begin{equation}
  \label{eq:lnL}
  Q(\mu)  = \frac{(y_{\rm exp} - \mu)^2}{\sigma_{\rm exp}^2}  +
  \left( 1 + \frac{1}{2 r_{\rm SM}^2} \right)
  \ln \left[ 1 + 2 r_{\rm SM}^2 
      \frac{(y_{\rm SM} - \mu)^2}{v_{\rm SM}}  \right] \;.
\end{equation} 

\noindent If the variance of $y_{\rm SM}$ is estimated very accurately
then $r_{\rm SM} \ll 1$ and by expanding the logarithm one recovers
the usual quadratic constraint.

As discussed in Ref.~\cite{bib:Cowan2019}, a usual least-squares
average with known uncertainties is equivalent to having Gaussian
distributed inputs.  As the tails of a Gaussian decrease very rapidly,
a Gaussian-distributed value is extremely unlikely to depart from its
mean by, say, five standard deviations.  The gamma variance model is
equivalent to replacement of the Gaussian by a Student's $t$
distribution (see, e.g., Ref.~\cite{bib:Cowan1998}), where the number
of degrees of freedom is $\nu = 1/2r^2$.  Thus a greater relative
uncertainty on an assigned systematic error corresponds to a lower
number of degrees of freedom and therefore to tails that are longer
than those of a Gaussian.

By minimizing the function $Q(\mu)$ of Eq.~(\ref{eq:lnL}) with
respect to $\mu$ one obtains the maximum-likelihood estimator
$\hat{\mu}$.  This is a weighted average of the SM prediction and the
experimental measurement, and in the current problem it is not of
direct interest.  The important result is rather the goodness of fit,
which can be quantified by the statistic $q = Q(\hat{\mu})$.

A higher value of $q$ represents increasing incompatibility
between the measured and predicted values $y_{\rm exp}$ and $y_{\rm SM}$.
This can be quantified with the $p$-value

\begin{equation}
  \label{eq:pval}
  p = \int_{q_{\rm obs}}^{\infty} f(q) \, dq \;,
\end{equation}

\noindent where $f(q)$ is the probability density function (pdf) of
$q$ and $q_{\rm obs}$ is its observed value.  For the usual
least-squares case, this would be a chi-square distribution for one
degree of freedom.  Here, however, because of the logarithmic term in
Eq.~(\ref{eq:lnL}), the pdf of $q$ is found to depart from chi-square
when the error-on-error parameter $r_{\rm SM}$ increases.

By construction the pdf of $q$ is independent of $\mu$, but as $r_{\rm
  SM}$ increases, $f(q)$ acquires a dependence on $\sigma_{\rm SM}^2$.
To find the desired $p$-value of the composite hypothesis, one should
take the maximum $p$ found for any $\sigma_{\rm SM}^2$.  Here this is
approximated by computing the pdf of $q$ with the parameters set to
their maximum likelihood estimators (MLEs) $\hat{\mu}$ and
$\widehat{\sigma^2}_{\rm SM}$ (see, e.g., Refs.~\cite{bib:CranmerPC},
\cite{bib:PDG} Sec.~40.3.2.1).  After $\hat{\mu}$ is found by
  minimizing $Q(\mu)$ from Eq.~(\ref{eq:lnL}), the corresponding
  estimator for $\sigma_{\rm SM}^2$ is given by

\begin{equation}
  \label{eq:sigmamle}
  \widehat{\sigma^2}_{\rm SM} = \frac{v_{\rm SM} + 2 r_{\rm SM}^2
    (y_{\rm SM} - \hat{\mu})^2}{1 + 2 r_{\rm SM}^2} \;.
\end{equation}

To find the $p$-value as a function of $r_{\rm SM}$, one can thus
generate values of $q$ with Monte Carlo using the MLEs from the real
data for $\mu$ and $\sigma_{\rm SM}^2$.  To determine a very small
$p$-value corresponding to a significance of $5\sigma$ or higher,
however, the required amount of simulated data becomes very large.  In
Ref.~\cite{bib:Cowan2019} it is shown that by using a simple
correction due to Bartlett \cite{bib:Bartlett}, one may obtain
$p$-values with far less computation.  This correction was used in the
present analysis and found to agree well with the full Monte Carlo
method in regions where the latter is computationally feasible.

As is the usual practice in Particle Physics, the $p$-value is
converted into an equivalent significance $Z$ according to

\begin{equation}
  \label{eq:significance}
  Z = \Phi^{-1}(1 - p/2) \;,
\end{equation}

\noindent where $\Phi^{-1}$ is the standard Gaussian quantile (inverse
of the standard Gaussian cumulative distribution).  The formula used
here is appropriate for a two-sided hypothesis test, i.e., a positive
or negative difference between measurement and prediction is regarded
as equally discrepant.  In the limit where $r_{\rm SM} \ll 1$ and the
logarithmic term in Eq.~(\ref{eq:lnL}) becomes quadratic, the
significance is given by $Z = \sqrt{q}$, i.e., the square root of the
minimized chi-squared.

\section{Significance of the muon $g-2$ anomaly}
\label{sec:results}

In this section the procedure outlined above is applied to the
significance of the muon $g-2$ anomaly.  It is up to the theory
community and in particular the authors of Ref.~\cite{bib:SMpred} to
assess the reliability of the error estimate for the SM prediction,
and therefore the significance of the observed discrepancy is
presented below as a function of $r_{\rm SM}$.  It is not intended to
imply here that the assigned uncertainty of 0.43 is incorrect, only
that it could be uncertain and thus one may ask what impact such an
uncertainty could have on the significance of the discrepancy.

\begin{figure}
  \begin{minipage}[c]{0.62\textwidth}
    \includegraphics[width=\textwidth]{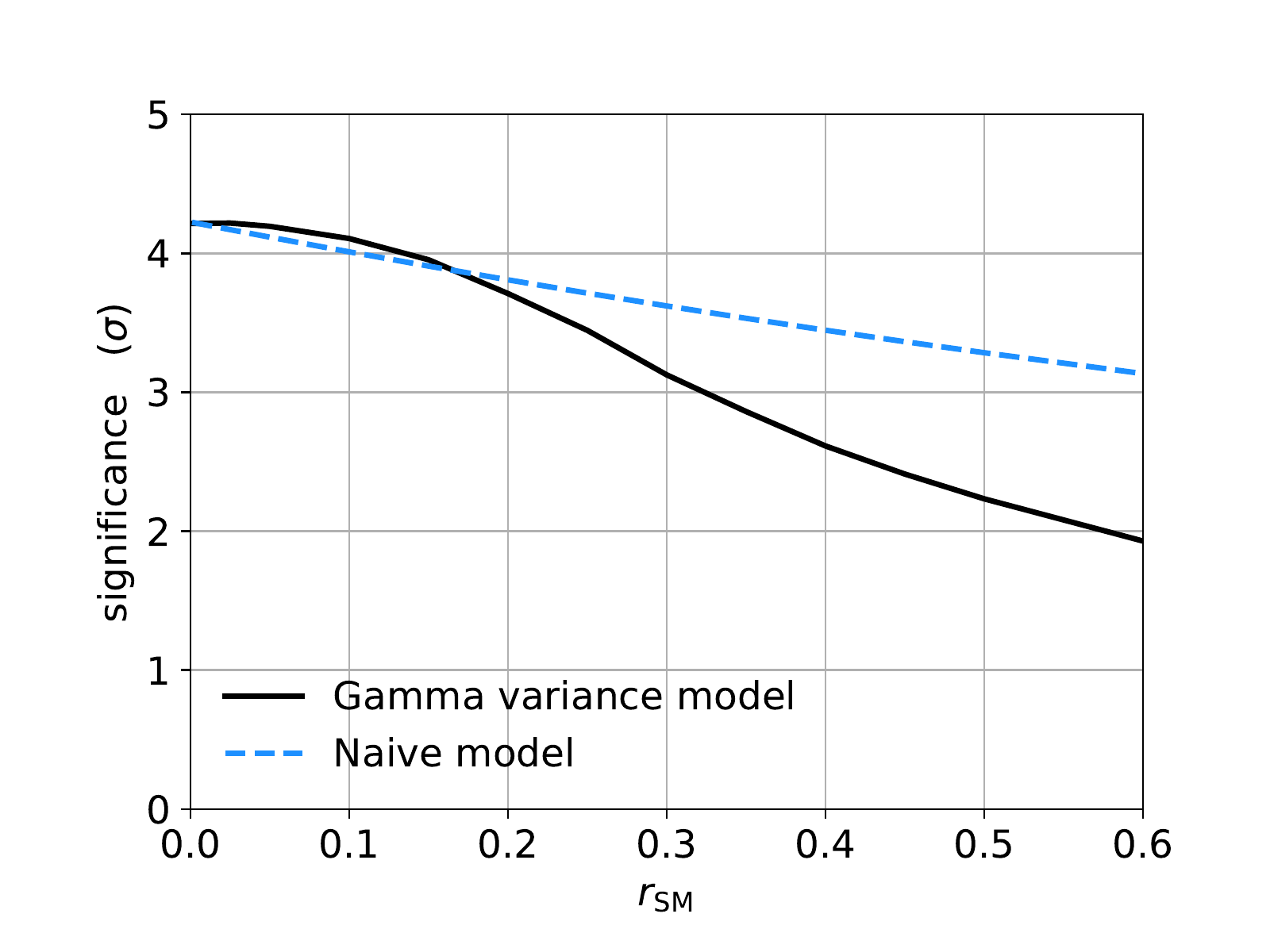}
  \end{minipage}\hfill
  \begin{minipage}[c]{0.35\textwidth}
    \caption{
      \small The significance of the muon $g-2$
          discrepancy as a function of the relative uncertainty
          $r_{\rm SM}$ in the quoted systematic uncertainty of the SM
          prediction (the relative ``error on the error'').
    } \label{fig:Z_vs_r}
  \end{minipage}
\end{figure}

Figure~\ref{fig:Z_vs_r} shows the significance in $\sigma$ of the
discrepancy between the experimental and predicted values as a
function of $r_{\rm SM}$ using the gamma-variance model described
above.   The figure also shows the significance that one would obtain
from a naive model, in which one simply inflates the SM uncertainty
$\sigma_{\rm SM}$ by

\begin{equation}
  \label{eq:naive}
  \sigma_{y_{\rm SM}} \rightarrow \sigma_{\rm SM} (1 + r_{\rm SM}) \;.
\end{equation}

\setlength{\unitlength}{1.0 cm}
\renewcommand{\baselinestretch}{0.9}
\renewcommand{\baselinestretch}{1}
\small\normalsize

\noindent As can be seen in the figure, the naive approach is quite
close to the gamma variance model for relative uncertainties below
around 20\% (i.e., $r_{\rm SM} \le 0.2$).  But already for $r_{\rm SM}
= 0.3$ the predictions differ substantially, with a significance of
3.12 from the gamma variance model and 3.63 from the naive model.  The
difference is increases at $r_{\rm SM} = 0.6$, with significances of
1.94 and 3.13.

\begin{figure}
  \begin{minipage}[c]{0.62\textwidth}
    \includegraphics[width=\textwidth]{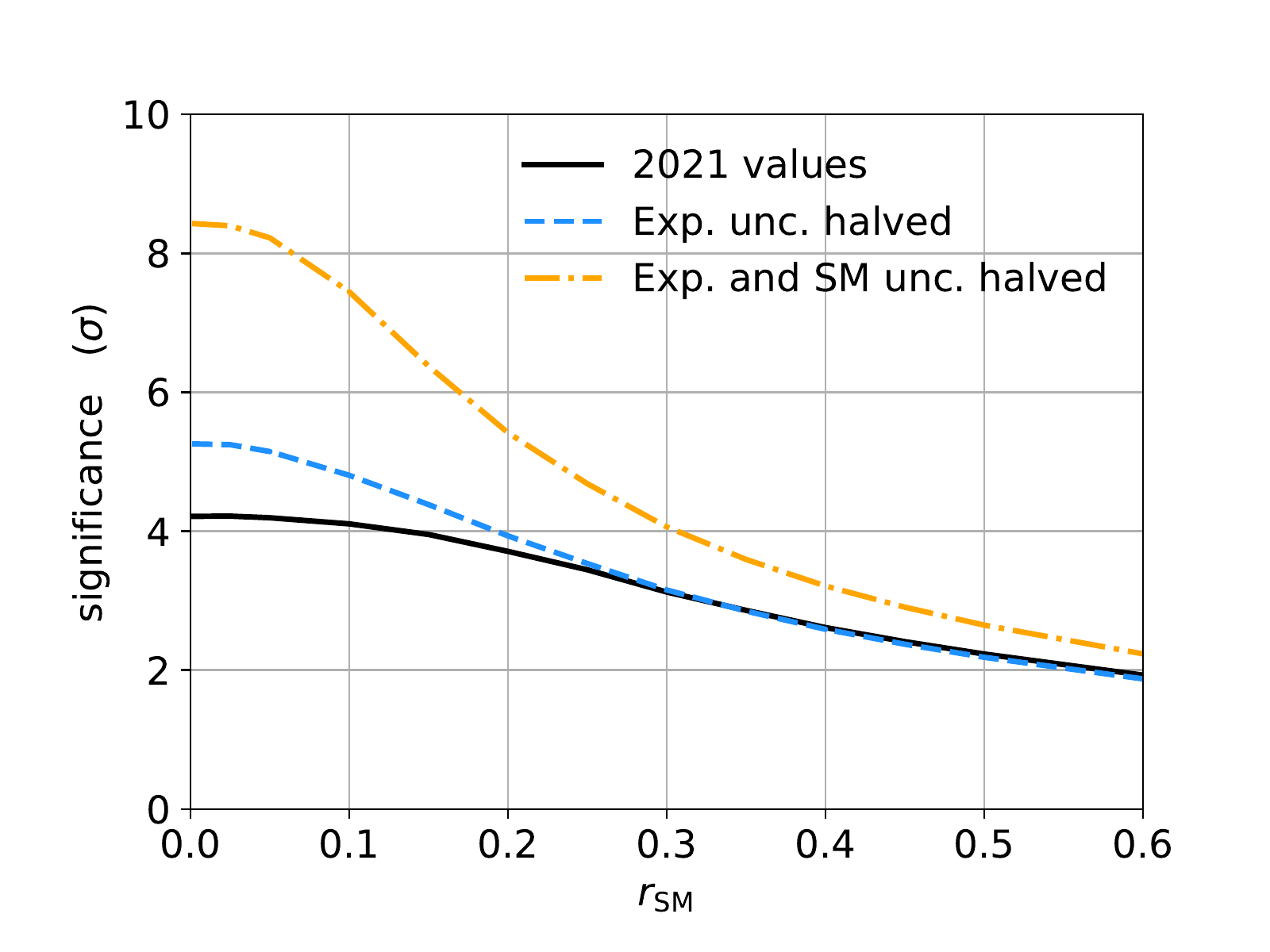}
  \end{minipage}\hfill
  \begin{minipage}[c]{0.35\textwidth}
    \caption{
      \small  The significance of the muon $g-2$
          discrepancy as a function of 
          $r_{\rm SM}$ assuming the 2021
          uncertainties (with SM uncertainty from \cite{bib:SMpred}),
          with the experimental
          uncertainty reduced by a factor of two,
          and with the SM uncertainty
          also halved.
    } \label{fig:Z_vs_r_future}
  \end{minipage}
\end{figure}

Additional data taking from the Muon $g-2$ Experiment is expected
reduce the experimental uncertainty.  Suppose that the measured value
remains the same but with half of its current uncertainty, i.e.,
$y_{\rm future} = 20.61 \pm 0.205$, and that the SM prediction remains
$y_{\rm SM} = 20.40 \pm 0.43$.  Figure~\ref{fig:Z_vs_r_future} shows
the significance of the discrepancy as a function of the relative
uncertainty on the error assigned to the SM prediction $r_{\rm SM}$.

\setlength{\unitlength}{1.0 cm}
\renewcommand{\baselinestretch}{0.9}
\renewcommand{\baselinestretch}{1}
\small\normalsize

From Fig.~\ref{fig:Z_vs_r_future} one sees that improved experimental
accuracy is of almost no benefit unless the accuracy with which one
assigns the SM uncertainty is kept below around 20\% Also shown on the
plot is the significance using an experimental uncertainty reduced by
a factor of two and an SM uncertainty $\sigma_{\rm SM}$ halved from
0.43 to 0.215.  The situation is improved somewhat by a reduction in
the SM uncertainty, but still only if this uncertainty is itself
assigned to correspond accurately to one standard deviation.  If the
naive recipe $\sigma_{\rm SM} \rightarrow (1 + r_{\rm SM}) \sigma_{\rm
  SM}$ is applied to the case of halved uncertainties one finds a
significance of $6.6\sigma$ at $r_{\rm SM} = 0.5$, compared to only
$2.6\sigma$ from the gamma variance model.

\section{Discussion and conclusions}
\label{sec:conc}

It should not be a surprise that the significance of the discrepancy
between SM prediction and experiment decreases when one supposes an
additional source of uncertainty, namely, the ``error on the error''
represented by $r_{\rm SM}$.  What is important to note, however, is
that simply inflating the corresponding uncertainty by a factor of
$(1+r_{\rm SM})$ does not adequately reflect the decrease in
significance, as can seen by the curves in Fig.~\ref{fig:Z_vs_r}.
Rather, the gamma variance model, which treats the assigned
systematics as gamma-distributed estimates, results in a more rapid
degradation of the significance for a relative uncertainty greater
than a certain level, starting around 20\% for this problem.  If the
relative uncertainty is 30\%, then the significance drops to
$3.1\sigma$, and it goes below $2\sigma$ for a relative uncertainty of
60\%.

If the all of the nominal uncertainties are reduced by a factor of two
relative to their current values, the discovery significance under
assumption of $r_{\rm SM} = 0$ is $8.4\sigma$.  But this reduces to
$5\sigma$ assuming a relative uncertainty in the theory error itself
of 22\%, and to $4\sigma$ for 30\%.  These results reflect the
intrinsic difficulty in establishing a discovery at a high
significance level if the uncertainties themselves are not accurately
determined, and as a consequence the corresponding distributions
acquire non-Gaussian tails.  This underlines the importance of
establishing appropriate procedures for quantifying systematic
uncertainties both for the muon $g-2$ anomaly as well as for any
investigations seeking to discover new phenomena.

\section*{Acknowledgements}
\label{sec:ack}

Many thanks for useful comments are due to Olaf Behnke, Kyle Cranmer,
Louis Lyons and Veronique Boisvert.  This work was supported in part
by the U.K.\ Science and Technology Facilities Council.

\end{document}